# Characteristic properties of Planacon MCP-PMTs

**Yu.A. Melikyan,**[a,1] **M. Slupecki,**[b] **I.G. Bearden,**[c] **J.R. Crowley,**[d] **D.A. Finogeev,**[a] **E.J. Garcia-Solis,**[e] **A.E. Guard,**[d] **A.V. Harton,**[e] **V.A. Kaplin,**[f] **T.L. Karavicheva,**[a] **J.L. Klay,**[d] **I. Morgan,**[e] **I.V. Morozov,**[a] **D.V. Serebryakov,**[a] **R. Stempek,**[e] **W.H. Trzaska**[g] **and N.O. Vozniuk**[a,f]

[a] *Institute for Nuclear Research of the Russian Academy of Sciences, V-312, 60-letiya Oktyabrya prospect 7a, Moscow, 117312, Russia*
[b] *Helsinki Institute of Physics, P.O. Box 64, 00014 University of Helsinki, Finland*
[c] *Niels Bohr Institute, University of Copenhagen, Nørregade 10, 1017 Copenhagen, Denmark*
[d] *California Polytechnic State University, 1 Grand Avenue, San Luis Obispo, CA 93407, United States*
[e] *Chicago State University, 9501 S. King Drive, Chicago, IL 60628, United States*
[f] *National Research Nuclear University MEPhI (Moscow Engineering Physics Institute), Kashirskoe shosse 31, Moscow, 115409, Russia*
[g] *University of Jyväskylä, P.O. Box 35, FI-40351 Jyväskylä, Finland*

*E-mail:* ymelikyan@yandex.ru

ABSTRACT: A systematic investigation of Planacon MCP-PMTs was performed using 64 XP85002/FIT-Q photosensors. These devices are equipped with microchannel plates of reduced resistance. Results of a study of their gain stability over time and saturation level in terms of the average anode current are presented. This information allows one to determine the lower limit of the MCP resistance for stable Planacon operation. The spread of the electron multiplication characteristics for the entire production batch is also presented, indicating the remarkably low voltage requirements of these MCP-PMTs. Detection efficiency and noise characteristics, such as dark count rate and afterpulsing level, are also reviewed.

---

[1] Corresponding author

**Contents**

## 1. Introduction

Microchannel plate-based photomultiplier tubes (MCP-PMTs) are a special kind of photosensor characterized by very high radiation tolerance [1, 2], compactness, and low sensitivity to magnetic field [3]. These features make them a preferable solution for precise-timing of photon detection in high-luminosity accelerator-based experiments [4-9]. Very few past and present experiments have used MCP-PMTs in large numbers. Consequently, several points about the characteristic properties of the mass-produced devices require clarification:

- The voltage required to bias some of the recent MCP-PMT samples for a given gain decreased by 500-1500 V relative to the values reported previously [3]. Batch characterization of such PMTs may clarify how typical these values are for the MCP-PMTs mass-produced recently;
- Average anode current is claimed to be limited by the MCP strip current, thus correlated with the MCP resistance [10, 11]. The latter is bounded from below by the MCP self-heating phenomenon [12]. Study of the linearity limit of low-resistance MCP-PMTs production batch may clear up the correlation and help to find the optimal compromise between the two parameters;
- The intricate manufacturing process of the MCP-PMTs under UHV conditions results in a significant variation of the devices' photocathode quality and noise characteristics. Thus, evaluation of just a few individual MCP-PMT samples may not be representative of the final production batch.

As part of the ALICE upgrade in 2019-2021 [13], the Fast Interaction Trigger detector (FIT) capable of detecting the precise time and luminosity of relativistic ion collisions, has been added to ALICE [14]. Among other subsystems, FIT features two compact arrays of Cherenkov counters with a total of 52 units. Each unit uses a multianode Planacon MCP-PMT [15] to read out Cherenkov photons.

The FIT project uses Planacon XP85002/FIT-Q MCP-PMTs, a customized version of the commonly used and commercially available XP5012/A1-Q model. The main differences between the two device options featuring same bialkali photocathode are reduced MCP resistance and different wiring of the backplane. Both devices are equipped with standard 25 µm-pore MCPs without atomic layer deposition (ALD) treatment. With the customized Planacon version, we have avoided crosstalk-related problems, reduced the thickness, and increased the device's dynamic range. A detailed description of the implemented modifications and their justification can be found in [16]. Photographs of the Planacon XP85002/FIT-Q are shown in Fig.1. The backplane photo (Fig.1, right) reflects the grouping of 64 device anodes into four independent readout channels (quadrants) equipped with individual coaxial sockets.

Sixty-four photosensors have been produced and delivered to CERN in 9 separate batches in 2018 and 2019. Each device was characterized in a dedicated test bench. The bench testing results obtained and presented below provide an understanding of the aforementioned points relevant to the characteristic properties of the mass-produced Planacons.

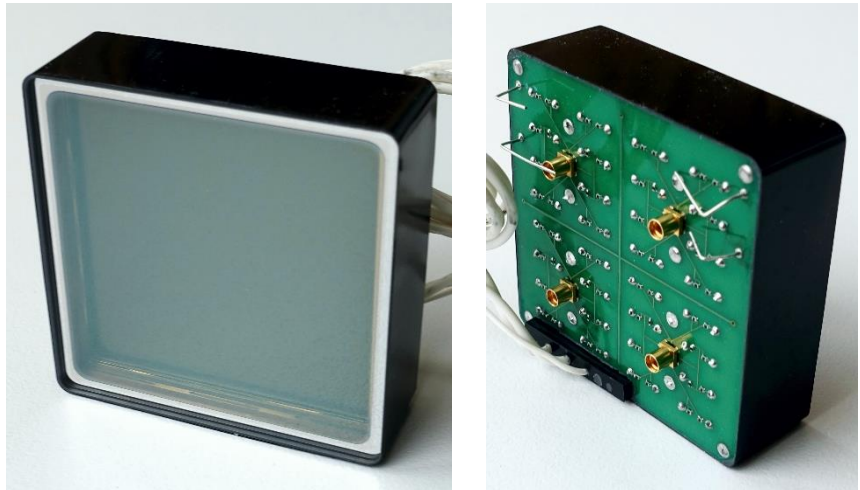

**Figure 1.** Front and back photos of Planacon XP85002/FIT-Q MCP-PMT as received from the factory. Dimensions of the sensor are 59x59x25 mm$^3$, sensitive area of a single channel – 26.5x26.5 mm$^2$.

## 2. The experimental setup and testing program

Our dedicated test bench with a capacity of up to 10 MCP-PMTs is shown schematically in Fig.2. The photosensors were placed in arrays of 5 units, each located in separate partitions of a custom light-tight box. Each array was illuminated by 440 nm pulsed light from a Picoquant PDL800-D laser with an LDH-P-C-440M laser head. The laser light was delivered to the photocathodes through a system of multimode optical fibres, splitters, attenuators (OZ Optics DD-100), and diffuse reflective screens. The latter were used to illuminate each of the PMT quadrants with equal light intensity. A reference PMT (Philips 56 AVP) was illuminated directly in a separate partition of the light-tight box.

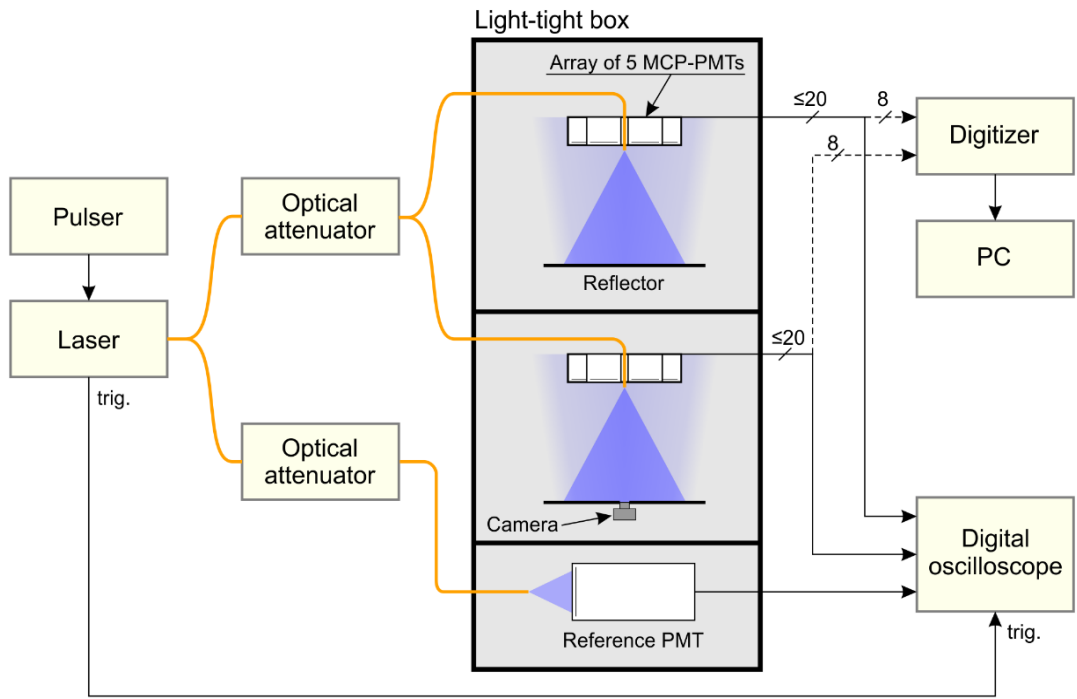

**Figure 2.** Simplified schematic diagram of the test bench.

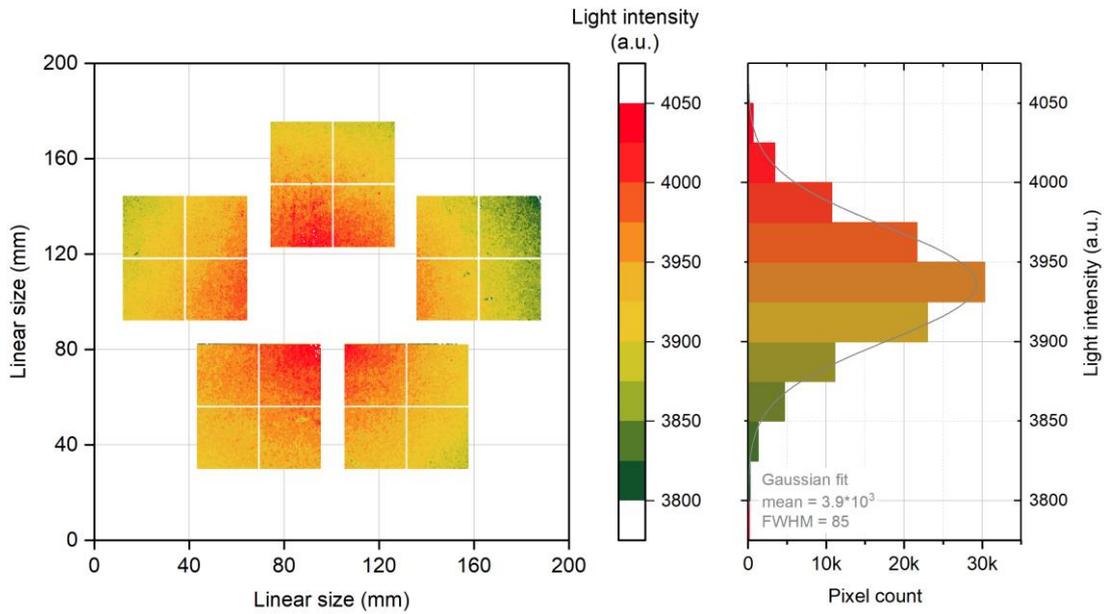

**Figure 3.** Distribution of light intensity at the entrance windows of MCP-PMTs in one array. The left graph indicates the exact location of MCP-PMTs within each array.

Figure 3 shows the distribution of light intensity across the array of 5 MCP-PMTs measured by a camera with a 12-bit CMOS sensor. For this measurement, the PMT array was replaced by another diffuse reflective screen for the proper light detection by the camera located opposite. As shown in Fig.3 (right), light intensity distribution across the sensitive area of the MCP-PMTs in the array exhibits a normal distribution with a width of 2% only (FWHM/mean).

The signal readout is performed with a 1 GHz 10 GS/s LeCroy WaveRunner 8104 digital oscilloscope and a 16-channel 5 GS/s CAEN DT5742 digitizer. The different readout options were used depending on the parameter measured as indicated in Table 1. Several parameters, such as the absolute gain and detective quantum efficiency (DQE), were measured for each quadrant. Other parameters were measured for a single quadrant only.

Table 1. List of parameters studied for each device indicating the readout option used.

| Parameters | Readout method |
| --- | --- |
| Absolute gain | Oscilloscope |
| Warm-up parameters | Digitizer |
| Linearity limit in terms of the average anode current | Oscilloscope |
| Level of afterpulsing | Oscilloscope & digitizer |
| Detective quantum efficiency | Digitizer |
| High voltage (HV) stability over time (>500 hours) | No readout, HV source logging only |

## 3. Absolute gain calibration

Proper operation of a detector equipped with many vacuum PMTs requires their gain-matching. Moreover, in contrast to PMTs with a classical dynode structure, MCP-PMTs can have a significant gain variation across their sensitive area. This problem can not be mitigated by gain-matching, requiring the application of different attenuation coefficients at the inputs of the readout electronics.

We have performed absolute electron gain calibration under pulsed illumination for all four channels of a single MCP-PMT at a time. The gain scan was started from a minimum voltage in the range 1100-1350 V (just below the recommended voltage for a $10^4$ gain). The illumination intensity was adjusted to $230 \pm 80$ photoelectrons per trigger to keep the pulse amplitude between 2 mV and 2 V in the gain region of interest: $10^4$-$10^6$. This ensured proper pulse digitization by the 8-bit oscilloscope ADC. With $7\cdot10^5$ MCP pores per quadrant, the probability of two-photoelectron detection by the same MCP pore was kept negligible. Laser light intensity was set constant throughout the measurement for each set of MCP-PMTs and monitored independently by the reference PMT, which was biased to a fixed voltage. The repetition rate was set to 1 kHz to mitigate the influence of the average anode current saturation [17, 18]. The gain scan was performed by increasing the bias voltage in 25 or 50 V steps. Each step was followed by a waiting period (≥5 min) before the actual measurement for a proper device warm-up.

The absolute value of the anode signal charge (Q) was determined by measuring the area of the output signal waveforms averaged over 2000 digitized samples at each bias voltage (U). Absolute electron gain $G_e$ was determined by measuring the single photoelectron (SPE) charge spectrum at the maximum voltage ($U_{max}$) at the end of the gain scan. For this measurement, light intensity was reduced so that a single photoelectron was detected in 10% of triggers only. According to the technique described in [19], PMT gain $G_e$ is equal to the difference between the mean charge of SPE ($Q_{SPE}$) and pedestal ($Q_{pedestal}$) peaks divided by the elementary charge $q_e$:

$$G_e(U_{max}) = \frac{Q_{SPE}(U_{max}) - Q_{pedestal}(U_{max})}{q_e} \quad (1)$$

Both values of $Q_{SPE}$ and $Q_{pedestal}$ were determined by fitting the SPE spectrum measured at the maximum voltage. Figure 4 shows a typical Planacon SPE spectrum fitted with a sum of Gaussian distributions accounting for the pedestal and SPE charge with an exponential distribution in between. For simplicity, the fitting function did not cover the multiple-photoelectron events constituting 0.5% of triggers, leading to an insignificant bias in the fit result.

According to the description of this fitting technique [19], the exponential distribution between the two gaussian peaks represents noise, covering the contribution of photon conversions inside the MCP pores, electron emissions from the dynode structures and under-amplified photoelectrons. Although, given the relatively large photon wavelength (440 nm) and relatively short time integration window (20 ns at 1 kHz rate), contribution of the first two phenomena to the SPE spectra measured is expected to be negligible. At the same time, contribution of the under-amplified photoelectrons may not be considered as "noise", but as an inherent feature of the electron multiplication process independent of the illumination conditions.

Assuming the entire exponential distribution is populated by the signals caused by the under-amplified photoelectrons, SPE charge can be defined as a weighted mean of this exponential part and the gaussian SPE peak, both corrected for the pedestal. Such alternative technique is described in [20]. Figure 5 presents the difference in the SPE charge as defined by the two techniques (A and B, described in [19] and [20] respectively). The typical discrepancy in SPE charge is 5.5%, resulting in the possibility of an equal error in the determination of gain values presented below.

To convert the anode signal charge, Q, measured at lower voltage values U to electron gain $G_e$, we have calculated the average number of photoelectrons ($N_{p.e.}$) detected by each quadrant per laser trigger:

$$N_{p.e.}(U_{max}) = \frac{Q(U_{max}) - Q_{pedestal}(U_{max})}{Q_{SPE}(U_{max}) - Q_{pedestal}(U_{max})} \quad (2)$$

A correction was introduced for the reference PMT signal amplitude ($A_{ref}$) to account for possible instabilities of the laser light intensity over the course of measurements. The final value of electron gain was calculated as follows:

$$G_e(U) = \frac{Q(U) \cdot A_{ref}(U_{max})}{N_{p.e.}(U_{max}) \cdot A_{ref}(U)} \quad (3)$$

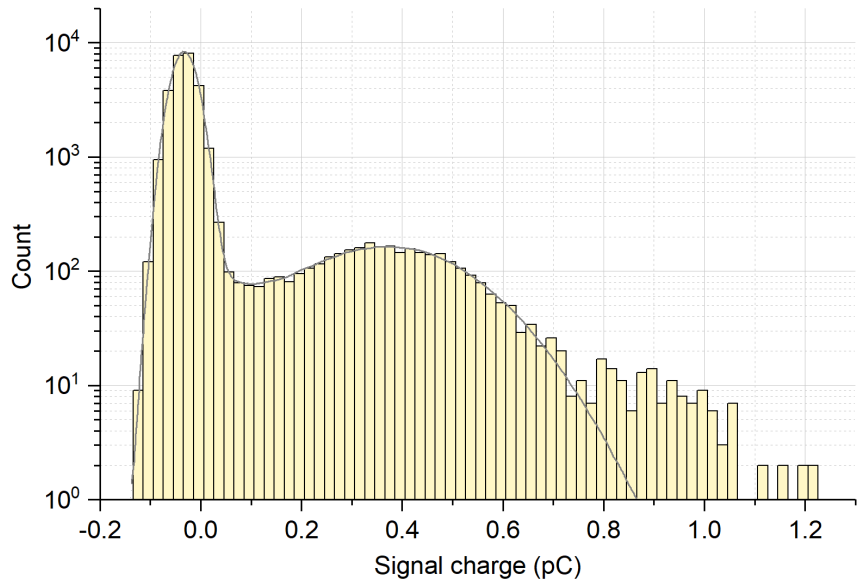

**Figure 4.** Typical SPE spectrum acquired from a single Planacon quadrant. In this measurement, each quadrant detected a photoelectron in ≈10% of triggers only. The fitting curve covers pedestal (mean = −0.03 pC) and single photoelectron peak (mean = 0.38 pC). The fitting curve does not account for the double-electron events constituting ≈0.5% of triggers.

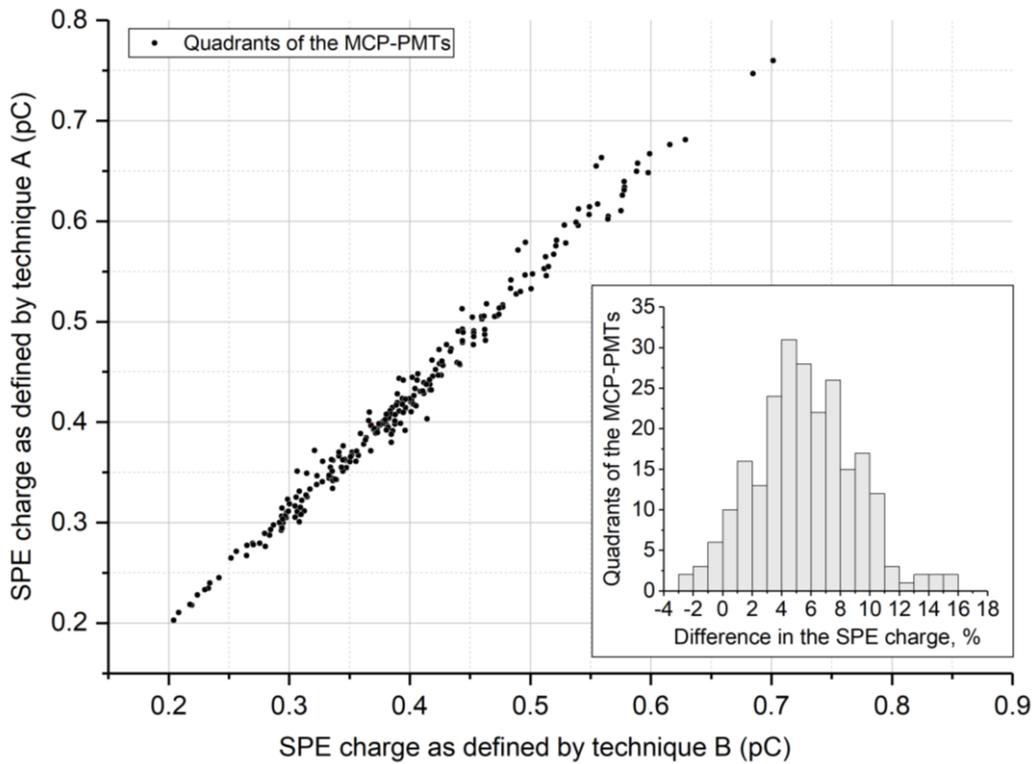

**Figure 5.** Difference in SPE charge as determined by techniques A [19] and B [20]. Technique A is used to derive MCP-PMT gain values presented below, resulting in a possibility of their overestimation by ~5.5%.

Sixty-four Planacon gain curves are shown in a single plot in Fig.6. The MCP-PMTs feature a significant spread of bias voltage required to achieve a given gain, though the shape of their gain curves is characteristic for any MCP-PMT [10]. Three sample curves are highlighted: two from the extremes of the distribution and one from its center. The curves do not follow an exponential law at electron gain values above ~$10^5$, where charge density inside the pores of the second MCP in a stack approaches the space-charge saturation limit. This limit is relatively low for the MCP-PMTs under study due to a direct coupling between the output of the first MCP and the input of the second one. Although, given the negligible probability of two-photoelectron detection by the same MCP pore, this phenomenon does not bias the electron gain values measured.

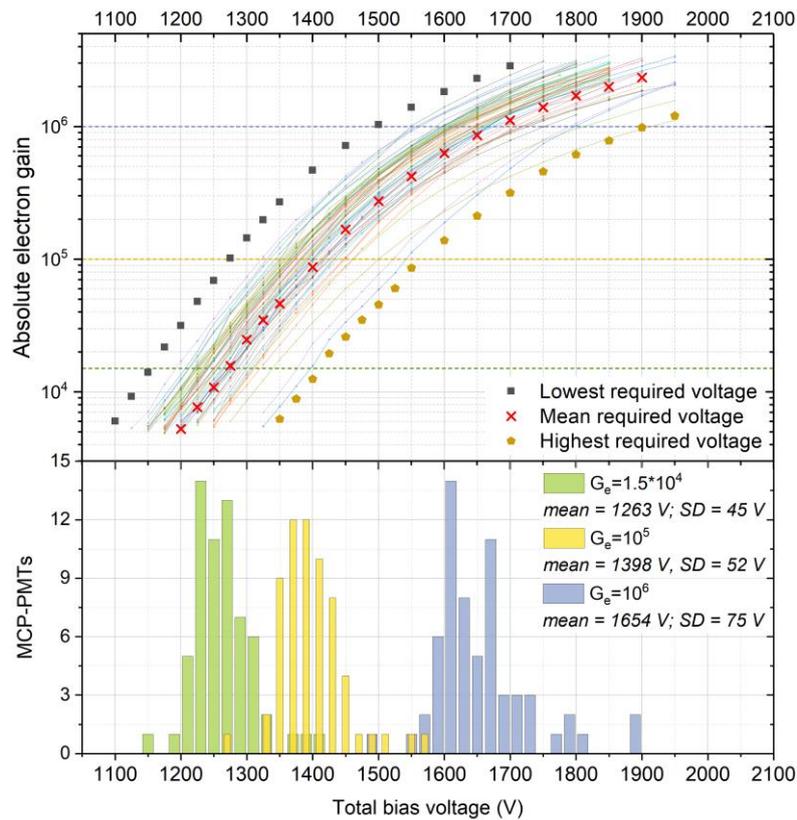

**Figure 6.** (Top) Electron gain versus bias voltage dependencies for the tested Planacon MCP-PMTs and (bottom) their projections for the gain values of $1.5 \cdot 10^4$, $10^5$, and $10^6$.

To reach the $1.5 \cdot 10^4$ electron gain default for FIT [17], the MCP-PMTs require bias voltage in the range from 1150 to 1410 V with the mean value of 1260 V (see Fig.6, bottom). The relatively low electric filed strength across the MCP may be beneficial to suppress the positive ion feedback and therefore to increase the lifetime of MCP-PMTs [21]. It is also beneficial in terms of HV discharge probability. Other similar non-ALD MCP-PMTs, including the older Planacon versions, require a significantly higher bias voltage for a given gain [22-30].

A comparison between the bias voltage required for $10^5$ electron gain and the same direct current (DC) gain is shown in Fig.7. Note that the former value is experimental, while the latter was listed in the specifications. At the factory, the gain is determined as the quotient between the anode current and the photo-current under constant device illumination. With such DC technique, one can measure the product of the MCP multiplication factor (so-called "electron gain") and collection efficiency (CE). Also, backscattered photoelectrons causing delayed pulses [31] are included in the device gain.

On an average, MCP-PMTs require 74 V higher bias voltage for $10^5$ DC gain than for the same electron gain. Or, being biased to a given electron gain, the tested MCP-PMTs feature ~45% lower DC gain. The difference arises from the limited CE of the MCP-PMTs, which we estimate to be ~45%. This is close to the 48% open area ratio (OAR) of the MCPs in the Planacon devices used in the FIT featuring a 25 μm pore diameter at 32 μm pitch.

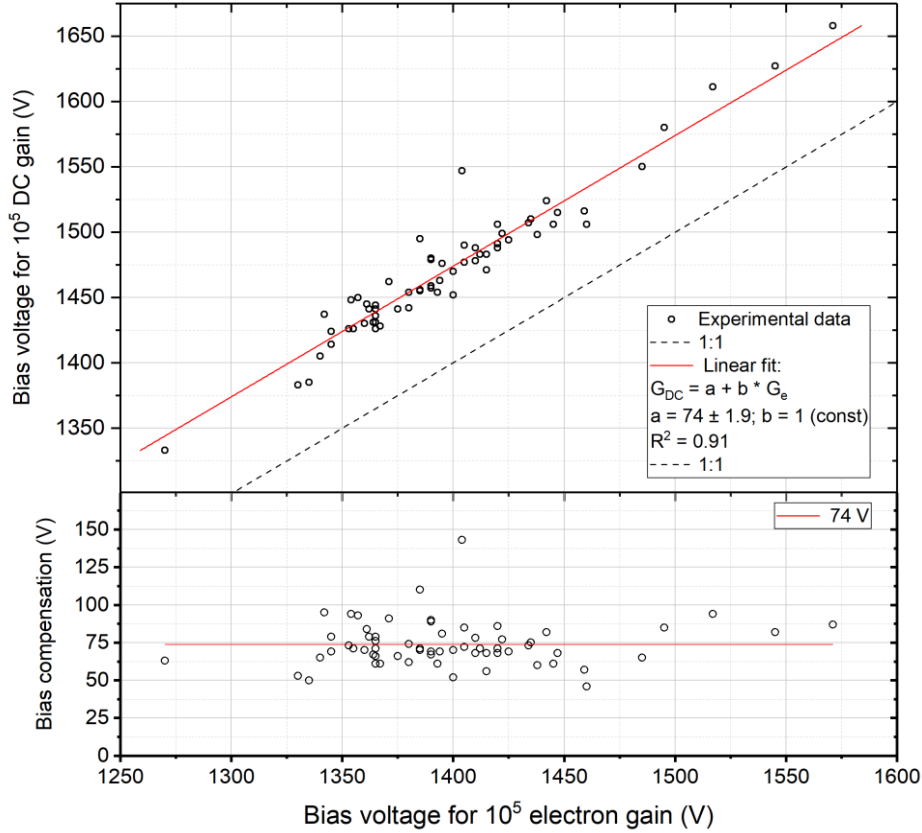

**Figure 7.** (Top) Bias voltage required for $10^5$ electron gain as determined experimentally and for the same DC gain as listed in the specs. The 1:1 line reflects the distribution expected for the case of 100% collection efficiency. (Bottom) Difference between bias voltage for $10^5$ DC gain and electron gain as a function of the latter.

## 4. Warm-up parameters and linearity

High-load operation of MCP-PMTs may require an enlarged linearity range for the average anode current (AAC). For example, FIT Planacons should be capable of operating linearly up to an AAC of 0.7-7 μA depending on the device location relative to the beam pipe [17, 18]. Standard Planacons are equipped with chevron stacks of two MCPs with a total resistance of 30-70 MΩ and a nominal AAC operational limit of 3 μA only [32]. At the same time, the MCP-PMT linearity limit for AAC is believed to be proportional to the MCP strip current, which is higher for the low-resistance MCPs [10, 11]. Because of this, and based on our previous experience with prototype MCP-PMTs of various resistance [18, 17], we have customized the FIT Planacons by limiting their stabilized MCP stack resistance to the range 12 MΩ ≤ $R_{MCP}$ ≤ 22 MΩ. With these limitations, the manufacturer declared an increased AAC operational limit for XP85002/FIT-Q MCP-PMTs of 10 μA [33].

The main disadvantage of using the low-resistance MCPs is increased resistive heating inside the PMT housing, which is an extremely poor heat sink. This effect is known to cause further reduction of the MCP resistance [12], limiting the possibilities for using very-low resistance MCPs. The FIT MCP-PMTs are biased through a standard voltage divider chain [33] with a single power supply. Under these conditions, the decrease in the MCP resistance leads to a reduction in the device gain. This effect is particularly pronounced right after the voltage ramp-up, implying a specific warm-up time to achieve a stable MCP gain.

We have performed a particular test to measure the Planacon's warm-up time. The tested devices and the reference PMT were illuminated with pulsed light at a 100 Hz rate, and their pulse amplitudes were measured at each trigger. Before the test, each device was stored unbiased for at least 12 hours, while the reference PMT was biased for at least 1 hour for a proper warm-up. Moreover, the instantaneous values of the bias current and voltage were logged by the CAEN N1470 power supply used to bias the Planacons.

Figure 8 presents trends for MCP stack resistance (a) and signal amplitude (b) measured for the first batch of four Planacons for FIT. The MCP resistance values were recalculated from the measured bias current and voltage values and known values of the voltage divider circuit. Each signal amplitude data point was calculated by averaging $6 \cdot 10^3$ measured values taken over each minute of measurement.

The dependence of signal amplitude on the MCP-PMT operation time was fitted with an exponential function to derive τ, the time for reducing of the pulse amplitude excess by a factor of $e$. The parameters measured for these four sample curves are summarized in the inset table. As can be seen, right after being biased to $U_b$=1.4 kV, the resistance of the MCP stacks in these devices was increased by 6% relative to the stable value. Similarly, a 3-6% excess in the device gain depending on the exact MCP stack resistance was observed. The overall distribution of the gain shift before the warm-up and the stabilization time τ is shown in Fig.10 for all FIT Planacons.

With MCP stack resistance ranging from 11 to 22 MΩ (Fig.9), the Planacons may require up to ≈30 minutes to decrease the pre-warm-up excess in gain to 2% (Fig.10). For our application, it is an insignificant side effect of the MCP stack resistance optimization to raise the device linearity limit. The linearity limit was measured by illuminating the tested MCP-PMTs and reference PMT by the pulsed light of a fixed intensity at a variable repetition rate from 20 Hz to 400 kHz. The entire photocathode was illuminated uniformly, although signals were only read out from a single quadrant of each MCP-PMT. The low-resistance base of the reference PMT ensured the linearity of its response within 2% for AAC of up to 2 μA at least. The light intensity at its entrance was attenuated to 0.2% of that delivered to each tested MCP-PMT to ensure its AAC does not reach this limit over the course of measurements even at high rates.

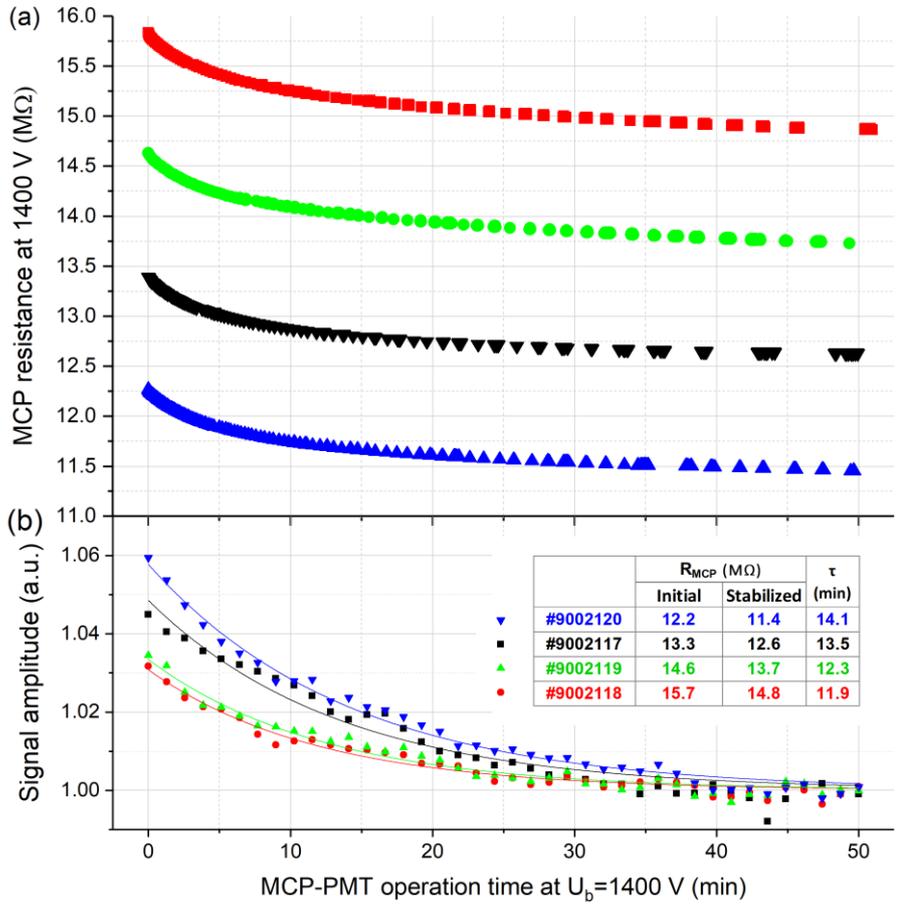

**Figure 8.** MCP stack resistance (a) and signal amplitude (b) trends during the warm-up of the first batch of XP85002/FIT-Q MCP-PMTs.

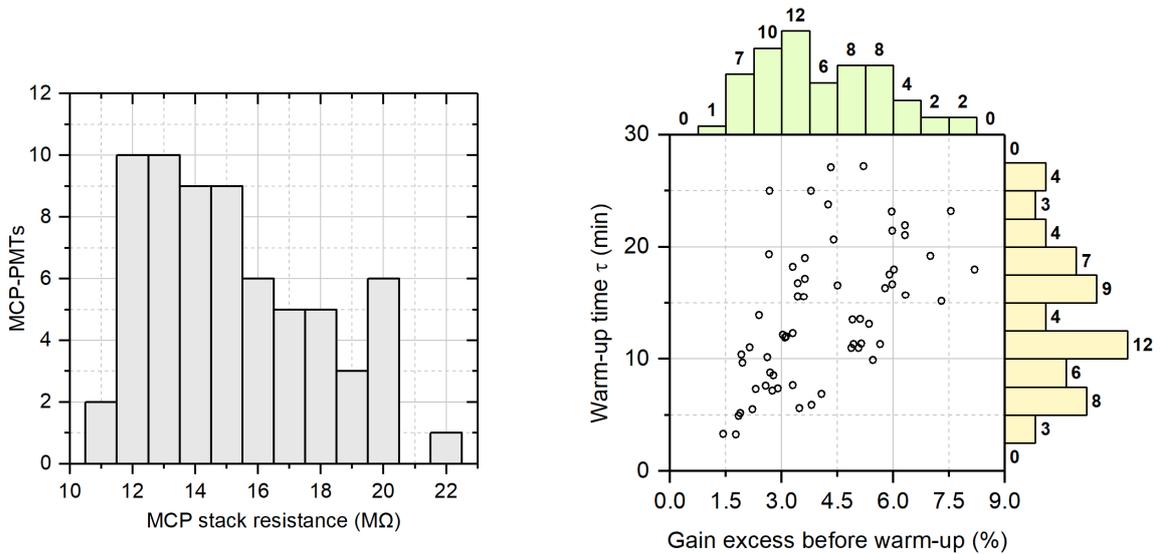

**Figure 9.** Stabilized MCP stack resistance of XP85002/FIT-Q Planacons measured after the warm-up.

**Figure 10.** Distribution of the warm-up parameters for Planacon MCP-PMTs with the MCP stack resistance range 11-22 MΩ.

Figure 11 presents a few sample saturation curves of a different shape measured for three FIT Planacons and the XP85012/A1-Q Planacon with MCP stack resistance of 35 MΩ. This set of sample curves demonstrates that the determination of the linearity limit is highly dependent on the linearity metric. For example, the AAC value at which the signal amplitude deviates by 1/3 varies by a factor of 13 among the four devices under discussion. Using a different linearity metric of 5% amplitude deviation reduces the spread to a factor of 2 only. Obviously, the main discrepancy comes from the increase in gain (or overlinearity [34, 35]) before the ultimate AAC limit is reached. Among the presented sample curves, this phenomenon is characteristic for Planacons #9002135, #9002129, and to a lesser extent for #9002028.

Among all FIT Planacons, #9002135 features the most significant increase in gain (30%) before the ultimate AAC saturation. For a consistent comparison of the tested devices, below we consider a 33% deviation in signal amplitude as the metric to determine the AAC linearity limit. The AAC linearity limit measured for each Planacon is presented as a function of MCP current (Fig.12) and overlinearity effect (Fig.13). Note that the total AAC is four times larger than the per-quadrant values.

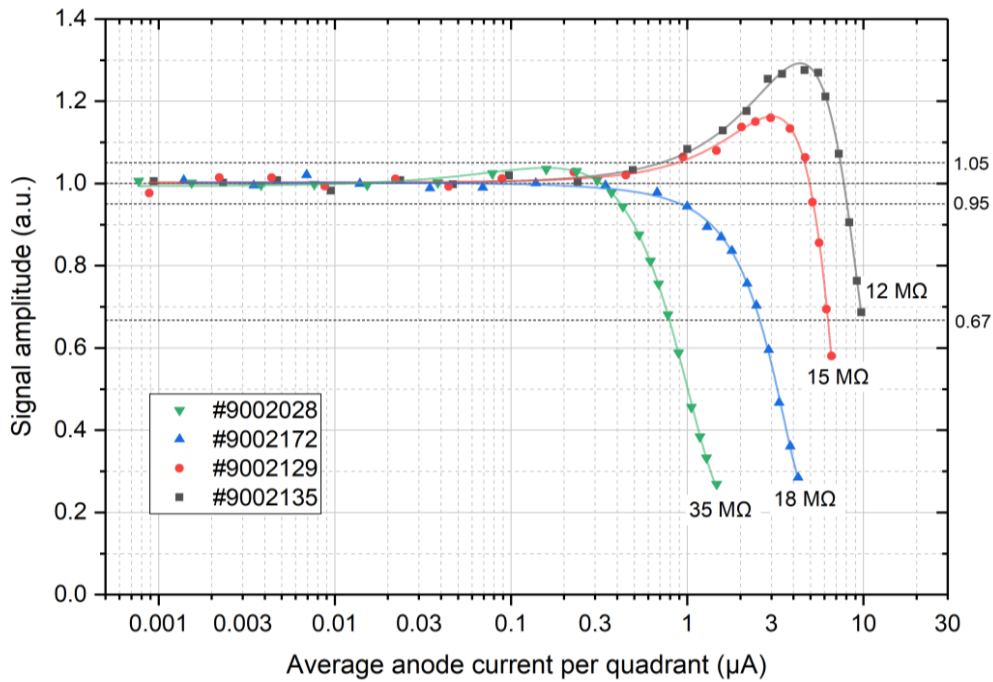

**Figure 11.** AAC saturation for four selected Planacon MCP-PMTs. AAC was increased by varying the illumination rate only.

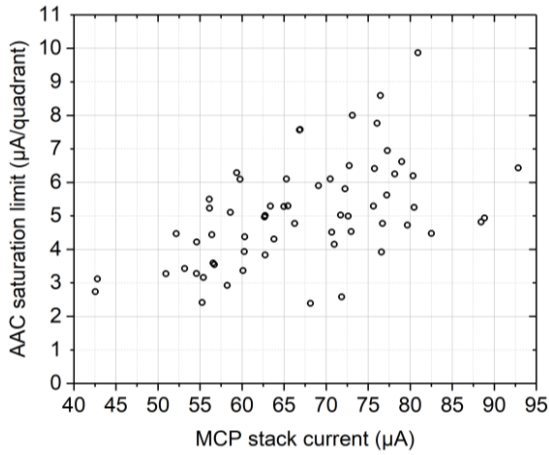 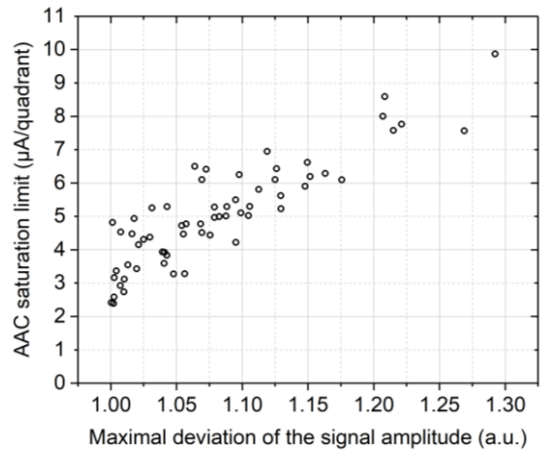

**Figure 12.** Planacon AAC linearity limit dependence on the MCP current.

**Figure 13.** Planacon AAC saturation limit dependence on the maximum increase of the signal amplitude (peak amplitude / initial amplitude).

A very moderate correlation is seen in the data presented in Fig.12 (Pearson's r = 0.52). Although, as seen from Fig.13, those devices with pronounced overlinearity exhibit higher linearity limit (Pearson's r = 0.88). This is an important conclusion illustrating the complexity of the manufacturing (and the purchase) of the MCP-PMTs of desired load capacity since selecting the MCP stack resistance and/or bias voltage for a given gain is not enough.

The conditions for the overlinearity manifestation are described in detail in [35, pp. 4-20...4-24] for the case of traditional dynode PMTs. These conditions are determined by the balance between the anode and bias currents corrected for the number of dynodes. A similar approach is complicated for the case of an MCP-PMT, since the average number of electron collisions with the pore walls ("number of dynodes") is not fixed, but depends on the bias voltage. Distribution of the average number of electron collisions along the amplification structure may be different even for two devices with identical measurable parameters, such as MCP stack resistance and voltage divider values. The reason lies in a possible mismatch between the resistance of two microchannel plates within the same stack. Ideally, their resistance should be equal, while the MCP manufacturer claims the resistance spread is ≤10% [36]. Although, since the output plane of the first MCP is directly coupled to the input plane of the second one, any external checking of this parameter is impossible. The imperfection of the resistance matching can cause a different contribution of the first and second MCPs to the device gain. As a result, predicting the linearity limit of a given MCP-PMT is extremely difficult without performing a direct measurement of the saturation curve.

Figure 14 shows the AAC limitations corresponding to a 5% gain deviation and 33% gain decrease for all tested devices. These parameters are correlated for MCP-PMTs having maximum overlinearity below 5% (blue data points in the bottom-right area of the graph). As could be seen from the presented distribution, 5% deviation from the linearity was observed at AAC in the range 0.5-3.7 µA per quadrant with a median value of 1 µA. These values correspond to the AAC of 2-15 µA per device and are even higher by a factor of 3 to 5 if using a 1/3 linearity metric. As mentioned above, in FIT, we expect AAC levels up to 0.2…1.8 µA/quadrant for different MCP-PMT slots across the FIT Cherenkov arrays. Only a quarter of the tested sensors preserve 5% gain stability at AAC above 1.8 µA/quadrant. However, by optimizing the location of the tested devices within the arrays based on the experimental results presented in Fig.14, we can meet the required detector performance.

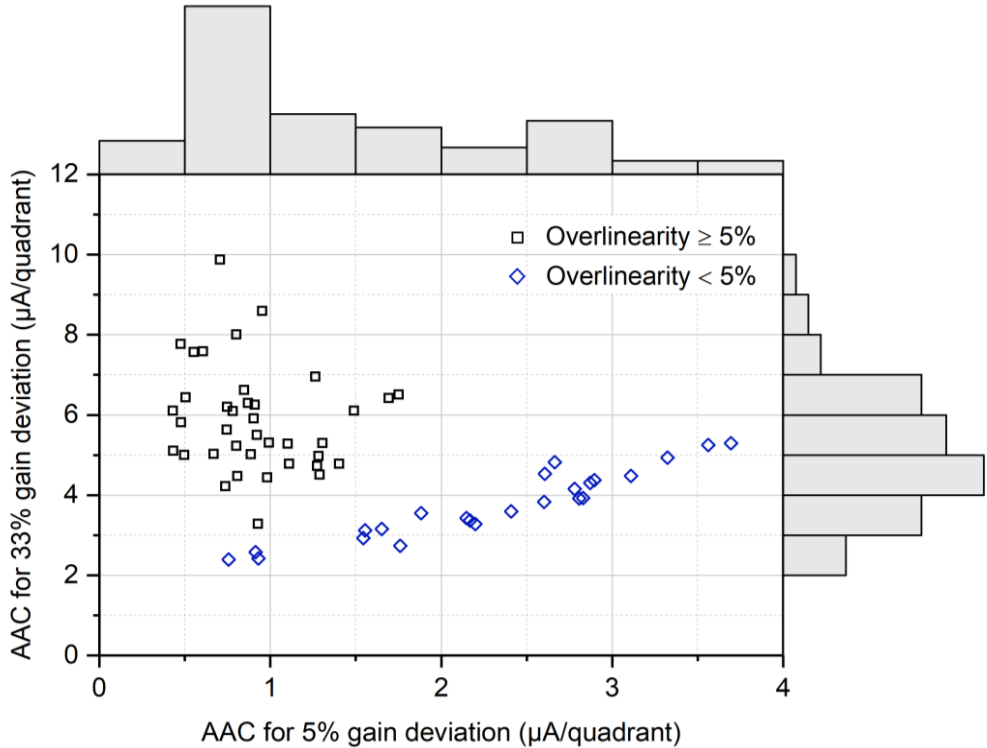

**Figure 14.** Dependence of the AAC limitation corresponding to 5% gain deviation and 33% gain decrease for all tested Planacons. Note that the total AAC limit per device is higher by a factor of 4.

## 5. Quantum efficiency for Cherenkov light

The manufacturer provides the efficiency of the Planacon photocathodes in terms of the luminous sensitivity for blue and white light (µA/lm). This information may be of limited relevance for the Planacon application to Cherenkov detectors since the majority of Cherenkov photons in solid radiators (e.g. quartz) are emitted in UV and VUV. Moreover, luminous sensitivity may be measured by the manufacturer only for a limited area within the photocathode.

We have measured the Cherenkov light yield from cosmic muons in the 2 mm-thick MCP-PMT quartz windows. Throughout the measurement, the MCP-PMTs were assembled in a telescope with their entrance windows directed upwards. Figure 15 presents the distribution of the amount of Cherenkov light detected in each tested quadrant, representing the spread of the relative detective quantum efficiency (DQE). To derive DQE, average signal charge from each quadrant was normalized to its absolute electron gain value measured before. Five quadrants featuring the lowest DQE (below 35 p.e.) are from two MCP-PMTs (#9002126 and #9002159 – see Table 2). Other quadrants of these PMTs feature twice this DQE. Since the PMTs are indivisible and equipped with a single HV input, equalizing the response of their channels is complicated (see Section 3). In the case of FIT, such photocathode nonuniformity is critical for the reasons of the insufficient photon statistics under 1 MIP detection by the outlier quadrants and the inevitable dynamic range reduction of the other quadrants in the attempts to equalize their response.

Table 2. Detected muon light yield in the 2 mm-thick Planacon entrance window for the two devices featuring the lowest DQE. Eventually, these two devices were replaced by the manufacturer.

| MCP-PMT | Detected muon light yield in the entrance window (p.e.) | | | |
|---|---|---|---|---|
| | q.1 | q.2 | q.3 | q.4 |
| 9002126 | 24±1 | 52±1 | 45±1.1 | 26±2 |
| 9002159 | 27±0.9 | 54±1.2 | 26±2 | 26±0.9 |

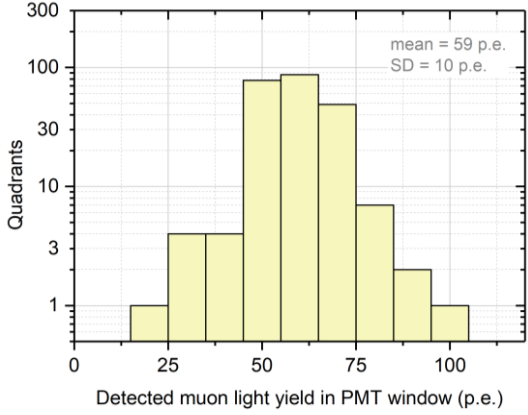

**Figure 15.** Distribution of cosmogenic muons' Cherenkov light yield in 2 mm-thick Planacon quartz windows detected by the individual quadrants in each MCP-PMT.

## 6. Noise characteristics

Afterpulses in MCP-PMTs could be caused by backscattered photoelectrons entering the first MCP and a positive ion feedback [37]. The mean charge of the afterpulses and the afterpulsing ratio [35] are good metrics to estimate the possible negative influence of such signals during detector operation. Moreover, potential deterioration of the vacuum quality in the MCP-PMT would be accompanied by an increase in these parameters.

We define afterpulses as signals above a 0.5 p.e. threshold arriving in a time window from 10 to 700 ns after the initial pulse caused by external illumination by the laser. Their number, $N_{690}$, should be corrected for the number of dark counts occurring typically in a random 690 ns window without any external illumination. We have estimated the dark count rate (DCR) by measuring the number of pulses $N_{290}$ detected in a time window from 300 ns to 10 ns before the initial pulse.

We define the afterpulsing ratio as the number of afterpulses per 300 p.e. of the initial pulse:

$$AR = \frac{N_{690} - \frac{(0.69 \cdot 10^{-6}\,s) \cdot N_{290}}{0.29 \cdot 10^{-6}\,s}}{N_{laser}} \bigg/ \frac{Q_{laser}/(q_e \cdot G_e)}{300\,p.e.}, \quad (4)$$

where $Q_{laser}$ and $N_{laser}$ are the charge and number of initial pulses caused by external illumination, respectively. $N_{laser}$ is also equal to the number of triggers.

The average charge of an afterpulse $q_{afterpulse}$, corrected for the average charge of a dark count, is the following:

$$q_{afterpulse} = q_{690} - q_{290} \cdot DCR \cdot (0.69 \cdot 10^{-6}\,s), \quad (5)$$

where

$$DCR = \frac{N_{290}}{(0.29 \cdot 10^{-6}\ s) \cdot N_{laser}},\qquad(6)$$

$q_{290}$ is an average charge of a signal detected from 300 to 10 ns prior to the initial laser pulse, and $q_{690}$ is an average charge of a signal detected from 10 to 700 ns after the initial pulse.

The noise parameters of the Planacons were measured using this technique at an electron gain of $G_e=10^6$ after keeping the MCP-PMTs in complete darkness for 12 hours. With the intensity of the initial pulse of the 440 nm laser below $10^3$ p.e., the probabilities of the superposition of afterpulses and photocathode excitation were kept negligible.

Typical noise characteristics measured are the following:
- Average afterpulse charge is distributed in the range 1.5-5 p.e. with a median value of 2 p.e.;
- The afterpulse ratio is distributed in the range 0.5-5 afterpulses per 300 p.e. with a median value of 1 afterpulse only.
- Among the tested sensors, only four devices feature a dark count rate above 25 kHz (ranging up to 250 kHz) at room temperature. The median DCR value for other devices is 3 kHz only.

For the case of FIT, the values reported above are significant only as a reference for future study in case of suspected vacuum leaks. By working at a $1.5 \cdot 10^4$ electron gain with ~100 p.e. detection threshold, no afterpulses are expected to be detected.

**Conclusions**

A systematic study of 64 Planacon XP85002/FIT-Q MCP-PMTs mass-produced for the ALICE FIT detector was performed. The manufacturer produced devices requiring 1260 ± 35 V to achieve the $1.5 \cdot 10^4$ electron gain. These values are remarkably low compared to other non-ALD MCP-PMTs reported in the literature.

The low bias voltage required is one of the reasons for the limited noise level of the devices. For the vast majority of the mass-produced devices tested at $10^6$ electron gain, only one afterpulse arises for each 300 p.e. of the initial pulse. The typical dark count rate is ~3 kHz only.

Contrary to the conclusions in earlier works of other authors [10, 11], only a moderate correlation was observed between the AAC saturation level and the MCP current for the narrow range of MCP resistance of the devices under study. At the same time, the influence of the internal parameters of the MCP stack, such as voltage distribution, is implicated. This complicates the manufacturing (and the purchase) of MCP-PMTs of the desired load capacity since selecting the MCP stack resistance and/or bias voltage is insufficient. It underlines the importance of the systematic characterization of the mass-produced devices. For our data sample, 5% deviation from the linearity was observed at AAC in the range 2-15 µA with a median value of 4 µA. The AAC for the saturation is higher by a factor of 3 to 5 if using a 1/3 linearity metric.

This relatively high saturation level was achieved for devices equipped with MCP chevron stacks of a resistance in the range 11-22 MΩ. A negligible side effect is that such devices can require up to 30 minutes warm-up period for a 2% gain stability over time.

Only two of the tested devices from the production series featured a twofold spread in the DQE of their quadrants reaching half of the DQE value averaged over the whole production batch. These devices were replaced by the manufacturer.

**Acknowledgments**

For Institute for Nuclear Research RAS, this material is based upon work supported by the Ministry of Science and Higher Education of the Russian Federation. For the University of Jyväskylä and HIP, the Academy of Finland grant is gratefully acknowledged. For California Polytechnic State University, this material is based upon work supported by the United States National Science Foundation under Grant Nos. PHY-1624988, PHY-1713894, and PHY-2012154. Any opinions, findings, and conclusions or recommendations expressed in this material are those of the author(s) and do not necessarily reflect the views of the National Science Foundation. For Chicago State University, this material is based upon work supported by the United States National Science Foundation under awards PHY-1613118, PHY-1625081, PHY-1719759, and HRD-1411219.